\documentclass[12pt]{article}

\usepackage{epsf}

\begin{document}

\centerline{\bf INFLUENCE OF A STRUCTURE ON SYSTEM'S DYNAMICS}
\centerline{\bf ON EXAMPLE OF BOOLEAN NETWORKS} 

\begin{center}
Olga V. Kirillova

Theoretical Physics Department, St.Petersburg State University,

St.Petersburg, 198904, Russia
\end{center}

\begin{abstract}

In this work we study the Boolean Networks of different geometric shape
and lattice organization.
It was revealed that no only a spatial shape but also type of lattice are
very important for definition of the structure-dynamics relation.
The regular structures do not give a critical regime in the investigated cases.
Hierarchy together with the irregular structure reveals characteristic
features
of criticality.

\vspace*{10pt}

\end{abstract}


\section{Introduction}		
\noindent

In 1969 S. Kauffman purposed to envisage a partial case of cellular
automata (CA) for investigation of biological processes \cite{Ka}. It
was found that subclass of CA called Boolean networks reveals a number of
properties successfully interpreted in terms of gene networks of living
organisms. After this the tool of Boolean networks became fairly useful
for investigations of a large range of phenomena: in spin
glass theory \cite{SpinGlas}, gene and neural networks
\cite{GNNet,Ka,Ka2,SS},
chaos controlling problem \cite{Chaos} etc.
Thus the study of properties of such systems acquired the undoubt
interest. It was found and investigated both numerically and analytically
many characteristic properties and regularities of Boolean networks
\cite{Ka,SpinGlas,BP}.

For random Boolean networks behavior it was revealed that there
are three different phases of system dynamics: chaotic, ordered and
marginal (which supposed to be critical) \cite{Ka,Ka2,SS}. It was pointed
out
that the behavior of such systems depends on many factors. Let us stop
at such characteristics of Boolean networks as its structure.

Investigations of the relation between a
structure and dynamics in real, for instance, biological systems are
extremely
difficult.
At least we want to understand one qualitativley.
For example, by the means of representation of gene regulation as a net of
units acting each other, which have only two possible states. In this case
the tool of Boolean Networks is very convenient.

So we envisage several networks of different forms with distinctions in
geometric shape as well as in action organization. Considering the
evolution of such systems for different dynamic rules, the choice of which
is resulted in obtained for random Boolean networks relations, which admit
to put system dynamics into ordered, chaotic or marginal class, we try
to understand how a structure influences on system behavior in different
phases. The investigation of system behavior has been made by the means of
analysis of measuring trajectories convergence in the phase space of
system's states, in other words through the measuring of the Hamming
distance ($Hd$) \cite{W}.

We do not try to perform any rule classification for cellular authomata
with a definit structure as it has been done in \cite{class}. Let us
stress that we investigate no only different lattice structures but
also different geometric shapes of the Boolean Networks.

Section 2 of this paper defines Boolean Networks and describes the
possible approach to its investigations. In section 3 the studied
structuries and rules of dynamics are presented. Section 4 describes 
obtained results. Finally, section 5 gives some conclusions.

\section{Definitions and approaches}
\noindent

Boolean Networks can be considered as an example of cellular automata
\cite{W,Tur}. The theory of cellular automata is quite
successfully
applied to study complex systems. \cite{ACA}

A Boolean network can well represents a big set of very different
structures of the system from random Boolean networks models introduced by Kauffman \cite{Ka}
with its
totally random structure to cellular automata with a regular lattice and
local
interactions.

A Boolean network is represented by a system of $N$ interconnected
elements with
only two possible states 0 and 1 (on/off). Any
   element in the system can has a connection with $K$
   elements, where $K$ can be varied for different units. Under term
{\it connection} it is assumed that $K$ other elements influence the center
element. According to a logical or Boolean rule every element is moved
to the next state. A state of the
system
   is defined as a pattern of states (on/off or 0/1) of all
   elements forming it. All elements are updated synchronously, moving the
system
  into the next state, and each state can has only single resultant one.
   The total space of the system's states is defined as all possible $2^N$
combinations of
   the elements' values in the system.
   Since the number of all possible states is limited and
   transition rules are fixed and do not depend on time, the system
   reaches a limit cycle or a fixed point called an attractor.
   Attractors may be envisaged as the "target area" of an organism,
   i.e. the cell's type in the end of development.
 Limit cycles can be considered in certain aspects as
   biological rhythms \cite{Ka,T}.

As was pointed out in studying of models on the basis of Boolean networks
\cite{BP,Ka}, the behavior of such systems can be varied in a wide range
from
order to chaos and in number of cases, has quite nontrivial character.

Eventually the choice of topology or a structure of interactions when
another parameters are fixed, one
can consider as one of the means of controlling chaos \cite{Chaos}.

Due to totally characterize behavior of the system, in other words to say
to which trajectory every from $2^N$ points of phase space belongs, one
need to get the phase portrait of space of states of the system, namely
the number of attractors, basins of attractors, a size of a limit cycle,
if a more detailed analysis is desirable the number (or percentage) of
Eden gardens, the number of trees, a transient length etc are calculated. If the system
size is big enough it is extremely difficult to do, therefore in notes about the
system behavior generally, it is reasonable to be limited by the more
rough characteristics as, for example, the Hamming distance.
Let us take one as the main characteristics of system behavior (in terms
chaos/order).

Hamming distance is a reasonable and often useful measure of distance in the
configuration
space of states of a binary system. It is defined as the number of bits
which are
different between the binary sequences $S_1$ and $S_2$. Usually normalized
Hamming distance ($Hd$) is considered
$$Hd=\frac{1}{N}\sum_{i=1}^N(\sigma_i(t)-\sigma_i'(t))^2$$
where
$N$ -- number of elements in the system,
$\sigma_i(t)$ -- the state of {\it i-th} element at the moment {\it t},
$S_1=\{\sigma_i\}_1^N, S_2=\{\sigma_i'\}_1^N$.

Any configuration corresponds to the point in the space of all possible
configurations. According to the rules of dynamics, in other words, system's
evolution,
each initial configuration traces out a trajectory in time. As it was
revealed if the process is chaotic, the trajectories of nearby
configurations diverge (in number of cases exponentially or as for instance, on 
lattices with discrete variables and nearest-neighbor interactions the
Hamming distance increases with a power law) in time, if the
system's dynamics is ordered then closed trajectories converge, in
the critical case
ones neither converge nor diverge, the distance between them preserves
almost the same in time.

Let us consider pairs of configurations which differ only in
a single site, i.e. initial $Hd$ is equal to $1/N$.

If we consider just two trajectories and on this basis make the conclusion
about system behavior in a whole, it seems to be no quite correct.
Thus due to get more adequate statistical picture we consider
statistically avereged Hamming
distance
defined by the next means.

At first let's take a state, where only in a random single position stay
the unit,

the next state is chosen so that difference from the former is only in
the first position and  calculate the convergence of these trajectories.

Further we calculate the convergence of the former trajectory and one, which
differ from it only in 3d position. And so far for all odd numbers.

So we investigate N/2 pairs of trajectories.

The next step. A state with couple of units in two
random positions is envisaged and then the previous step is repeated.

And so far for states with $3,4,...,N-1$ units.

Generically we have done uniform selection from the total space of states
and
investigated $\frac{N(N-2)}2$ pairs of trajectories.

In total we have the average $Hd(t)$ defined as
$$Hd(t)=\frac{2}{N^2(N-2)}\sum_{\sigma,\sigma'\subset \Omega} 
\sum_{i=1}^N(\sigma_i(t)-\sigma_i'(t))^2$$

where $\Omega$ is the set of trajectories choosen by suggested above means. 

\section{ Structures and rules of dynamics}
\noindent

We have considered seven two-dimensional structures with connectivity
$K=3$.

We tried to cover more
different structures both in the sense of a spatial shape (ribbon closed to
circle, torus, sphere, cone) and in  the sense of links' organization (regular lattices,
loops, cascades or hierarchy structures, feedback and autoregulation)
. In striving to answer the question, how the type of
lattice, the spatial organization, boundaries, autoregulation, hierarchy
influence the behavior of a system and how the structure and
dynamic's rules interact, we used two sorts of rules namely, 
homogeneous (the same for all
elements) and
heterogeneous (one for odd and another for even elements).
The choice of the rules in the model is reasoned by the fact revealed
early for random Boolean networks ones, 
that the average connectivity of a network and rules governing its
behavior
are related by the next formula
$ K_c=\frac{1}{2p(1-p)}$.
As it was obtained if
$\langle K\rangle<\frac{1}{2p(1-p)}$
(where
$\langle K\rangle$ -- the average connectivity,
$p$ -- the bias of the rule of dynamics)
dynamics have to be ordered,
if
$\langle K\rangle>\frac{1}{2p(1-p)}$
we have to get chaos
and if
$\langle K\rangle=\frac{1}{2p(1-p)}$ the behavior is critical
\cite{SS,C}.
We use the same rules for all investigated structures.

\begin{itemize}

\item
The first subject of our investigation is a ribbon closed to a circle. The
elements are arranged on the edges of the ribbon. Each of them has links
with its left
and right neighbors and with the element opposite to it on the other edge of the
ribbon.

\item
The second system has the structure
represented by the directed graph, namely the binary
tree where neighbors of an element are two successors for it nodes plus
autoregulation (i.e. an element is itself neighbor). The last layer
is closed in a circle. So we have the hierarchical model on 
a cone with autoregulation.

\item
In the third case we investigated the same but undirected graph.
Neighbors of an element are two successors for it nodes plus its
ancestor, in other words, here undirection excepts autoregulation.
Thus we have a cone without autoregulation.

\item
The fourth investigated structure is represented by the regular honeycomb 
lattice closed
in a torus.

\item
The fifth case contains the same as in the previous case lattice but here
the
boundaries' elements are selfregulated.

\item
In the sixth case we have the same lattice forming a sphere.

\item
The seventh case is represented by the regular squared lattice closed
in a torus where neighbors of an unit are the left and the right and one
above it. So we have a closed cascade of hierarchical regulation.
\end{itemize}

Our rules are:

{\bf a}. the homogeneous rule -- if all three element's neighbor are
switched on then
the element will be switched on and it will be switched off in any other case.
In the number presentation it is 01111111 ($p=0.125$) (Here and below 0 in the rule
means on and 1 means off. Each figure corresponds to the neighbor's state,
where ones
are situated from all 0 to all 1
\vskip-0.3cm
\begin{table}[htbp]
\tt 0 0 0 0 1 1 1 1
\\
\tt 0 0 1 1 0 0 1 1
\\
\tt 0 1 0 1 0 1 0 1 
\\
\tt - - - - - - - -
\\
\tt 0 1 1 1 1 1 1 1 
\end{table} 
\vskip-1cm
\hskip2.8cm
)

{\bf b}. here we have the same rule as in the point {\bf a} for even elements
and the
next for odd ones -- if all three element's neighbor are switched off then
the element will be switched on and it will be switched off in any other case.
In the number presentation it is the same as above plus 10000000 ($p=0.125$)

{\bf c}. an element will be switched on only in case both 1st and 2d its
neighbor or 3d have such state.
In the number presentation it is 01000010 ($p=0.25$)

{\bf d}. here we have the same rule as in the point {\bf c} for even elements
and for
odd ones -- if 1st or both 2d and 3d element's neighbor are switched on then
the element will be switched on and it will be switched off in any other case.
In the number presentation it is the same as above plus 00011000 ($p=0.25$)

{\bf e}. an element will be switched on only in case any pair of its
neighbors has such state.
In the number presentation it is 00010110 ($p=0.375$)

{\bf g}. here we have the same rule as in point {\bf e} for even elements
and
for odd ones -- if only a single of element's neighbor is switched on then
the element will be switched on and it will be switched off in any other case.
In the number presentation it is the same as above plus 01101000 ($p=0.375$)

\section{Results}
\noindent
We pay attention to the next characteristics :
percentage of convergence couple of trajectories in the phase space {\it 
pc}, maximum pattern
difference between them {\it mp} and
behavior $Hd(t)$ (it is defined as period of the function). The
data of percentage convergence are presented in table 1, maximum pattern difference
and character of behavior $Hd(t)$ -- in table 2
and table 3 correspondingly.

Let us make some explanatory notes to the table 3.
 After a finite number of steps the system comes to a stationary regime
that can be either a fixed point or a limit cycle. So when we follow
the evolution a pair of trajectories one can observe that after a number
of steps if the trajectories have not converged the Hamming distance is
constant or some periodic function of time, moreover the period can be very large,
that point out that at least one trajectory belongs to a large limit cycle,
that is usually proposed by a feature of chaos. It is true for each 
pair of trajectories, therefore the bigger the period of $Hd(t)$ is the more
chaotic system behavior supposed to be.
A number of iteration steps before trajectories have converged one can
consider as the maximum transient length.

As it is easy to see in a whole for different structures the results vary.
But in spite of this there is a general tendency in systems'
behavior and
several fluctuations (1{\bf b}, 3{\bf c}, 1{\bf d}, 7{\bf d}, 6{\bf g}).
For all {\bf a} and {\bf e} cases percentage of convergence is always
very high ($100 \pm 0.0004$ for {\bf a} case and $99.93 \pm 0.37$ for
{\bf e} one). Excepts second structure the maximum patterns difference is
always more than
19\% for {\bf a} case and it is always less than 1.5\% for {\bf e} one.
These cases correspond to the ordered behavior.

Qualitatively {\bf g} case always demonstrates chaotic behavior (too large period of
$Hd(t)$, quite low the convergence percentage). Except
second structure the maximum patterns difference is always more than 48.
At the same time
the character of $Hd(t)$ behavior is very variable.

Case {\bf b} has very low the maximum patterns
difference (less than 1\%) for all structures and the results are
close to each other. Majority of
structures in this case demonstrate rather chaotic behavior.

The system
behavior is analogous for all {\bf c} cases excepts third
structure.
One can put it into the category of chaos.

In {\bf d} cases besides the second and 7th structures we have ordered
behavior. 

Overall, in our case the obtained results do not confirm those
for random Boolean networks. These results allow to conclude that
heterogeneity/homogeneity of
rules influences on system's behavior in more extent than the $K,p$ relation.
In boundary cases ({\bf a} {\bf b} and {\bf e} {\bf g})
heterogeneity result in chaotic dynamics, homogeneity -- to order one and vice
versa for middle cases ({\bf c} {\bf d}). Let us stress that ordered dynamics in {\bf a}
and {\bf e} cases has not special situations.

Let's stop at some interesting results and distinctions.

The case 1{\bf g} has the minimum percentage of convergence.
Eventually in this case there are
big number of attractors, large pattern
difference between them. We stress that in spite of the high
convergence percentage 1{\bf d} case reveals obvious chaotic features, if
trajectories do not converge the difference between them becomes quite
big.
The character of $ Hd(t)$ is similar to 1{\bf g} case (Fig. 1), (Fig. 2).

\begin{figure}
\centerline{\epsfxsize12.0cm\epsffile{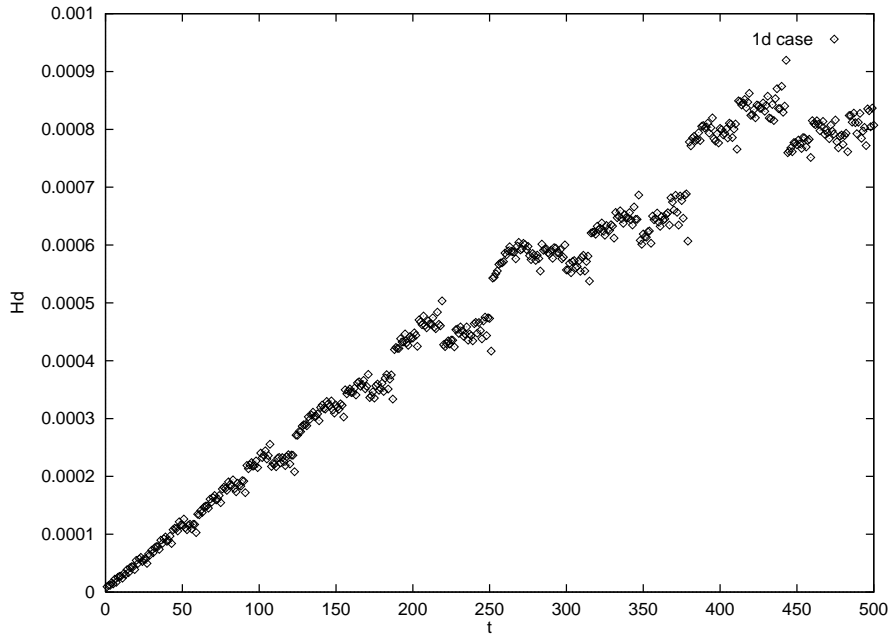}}
\caption{ $Hd(t)$ for the {\bf d} rule on the 1st structure. Size of
the system 1024.}
\end{figure}

\begin{figure}
\centerline{\epsfxsize12.0cm\epsffile{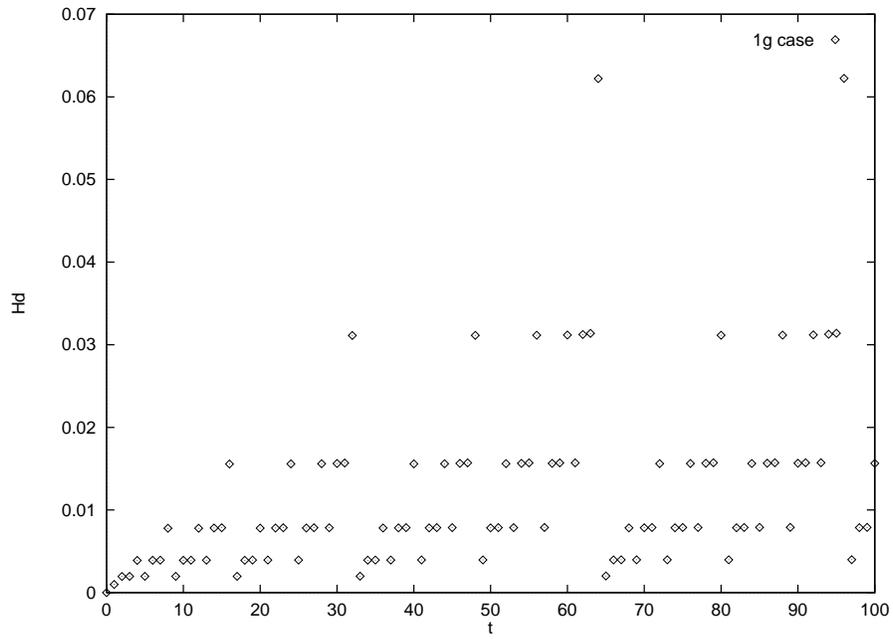}}
\caption{$Hd(t)$ for the {\bf g} rule on the 1st structure. Size of
the system 1024. }
\end{figure}

As for the second structure, here we observed some critical features in
behavior of the system, especially in the {\bf b} and {\bf d} cases.
For the {\bf g} case we obtain that
$Hd(t)$ has period equal to 8, percentage of convergence is
quite high for chaotic behavior (Fig. 3). Overall this structure has
no large maximum pattern differences, moreover it has the smallest value
for both {\bf a} and {\bf g} cases and the smallest average.

In case of the third structure we can see strongly different from all
other rules (from the same 
category) the result for the rule {\bf c}.
This case is characterized by too large (for rule {\bf c}) convergence percentage
and at the same time $Hd(t)$ has quite large period (equal 12) (Fig. 4).
Let us note that 3{\bf c}
case has the maximum pattern divergence equal to system's size. (It is the maximum among
all other cases.)
The minimum convergence percentage is observed not in {\bf g} but in {\bf b} case.
{\bf G} and {\bf a} cases have the maximum pattern difference and the maximum average
of this value (for its categories). In (Fig. 5) one can see  the $Hd(t)$
distribution for
{\bf
g} cases on
this structure.

\begin{figure}
\centerline{\epsfxsize11.5cm\epsffile{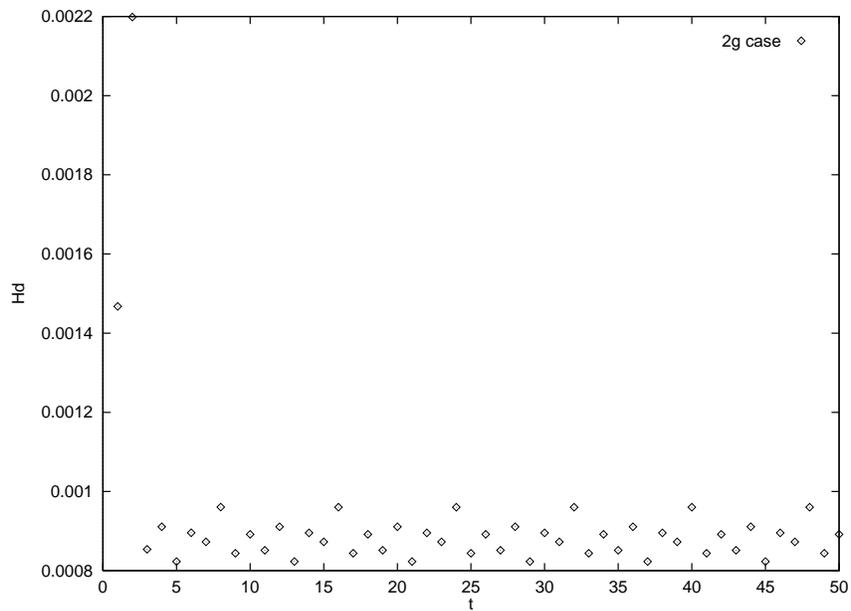}}
\caption{ $Hd(t)$ for the {\bf g} rule on the second structure. Size of the
system 1023.}
\end{figure}

\begin{figure}
\centerline{\epsfxsize11.5cm\epsffile{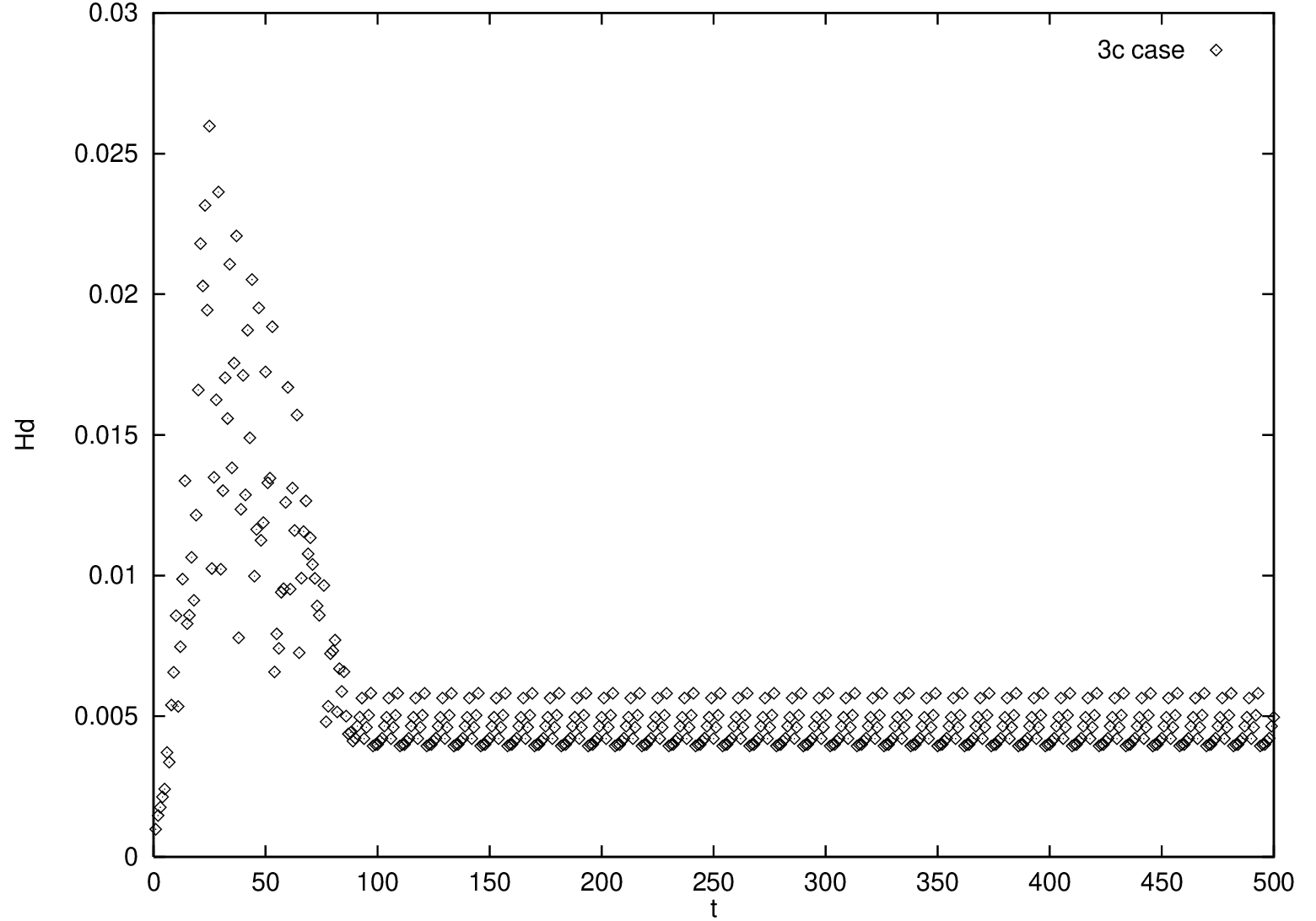}}
\caption{ $Hd(t)$ for the {\bf c} rule on the third structure. Size of the
system 1023.}
\end{figure}

\begin{figure}
\centerline{\epsfxsize11.5cm\epsffile{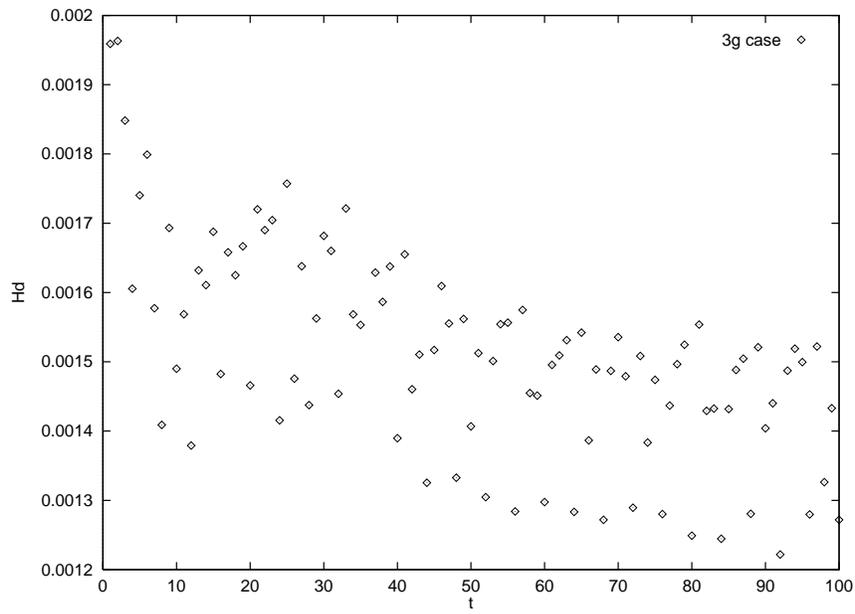}}
\caption{ $Hd(t)$ for the {\bf g} rule on the third structure. Size of the
system 1023.}
\end{figure}

Comparing 4th and 5th structures one can conclude that the influence of
boundaries is negligibly small as is the one of the autoregulation. System's
behavior is determined rather by the construction of the lattice. Here
the behavior is either
chaotic or ordered (not any critical features) with very
strong differences between them.
Excepting {\bf b} and {\bf g} cases the behavior on these structures is close to
the 1st one. In (Fig. 6) one can see the $Hd(t)$ distribution for {\bf g}
cases on these
structures.

\begin{figure}
\centerline{\epsfxsize11.5cm\epsffile{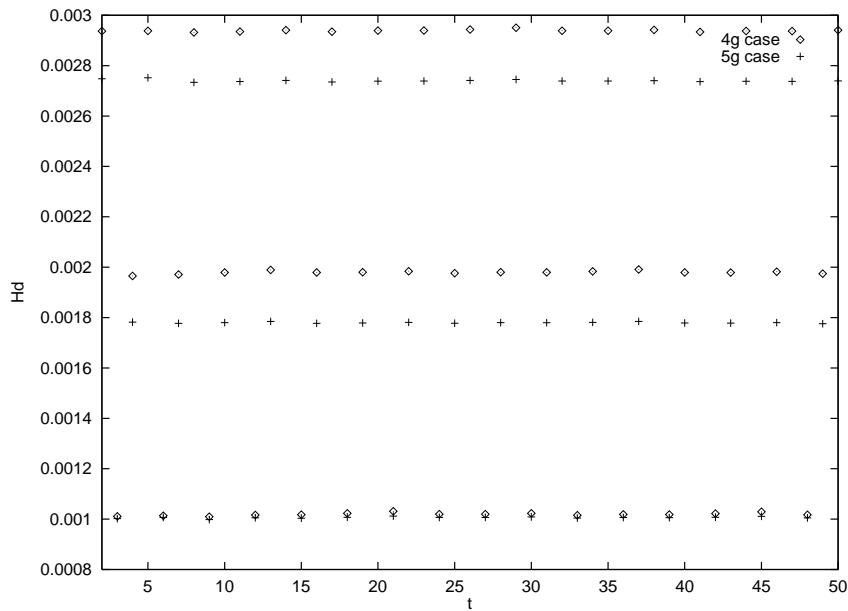}}
\caption{ $Hd(t)$ for the {\bf g} rule on the 4th (diamonds) and 5th (cross) structures.
Size of the systems 1200.}
\end{figure}

Behavior on the structure 7 is quite close to the 4th and the 5th ones excepting
{\bf d} case. This confirms a regular structure and a shape influence. The
7{\bf d} case strongly differ from other (from its category) by low convergence
percentage. In (Fig. 7) one can see $Hd(t)$ distribution for {\bf g} case
on this structure.

\begin{figure}
\centerline{\epsfxsize11.5cm\epsffile{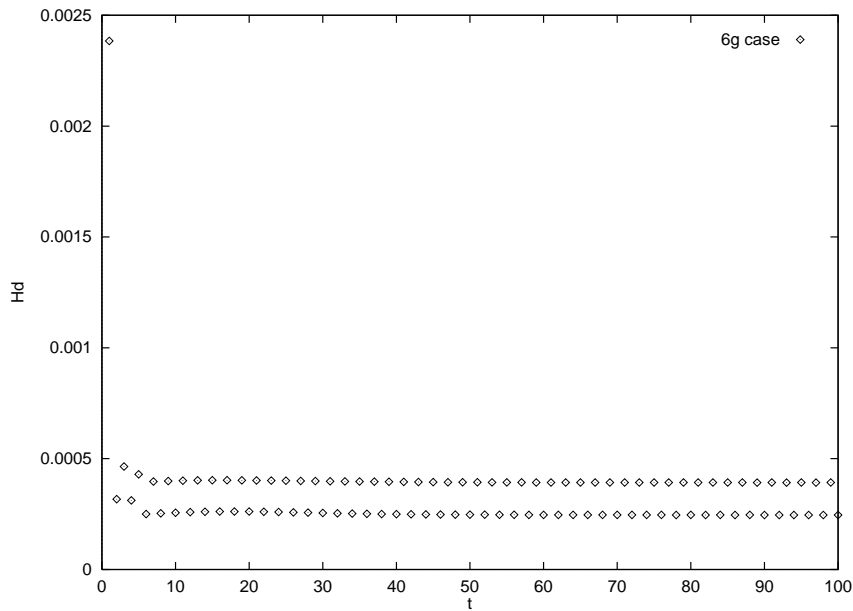}}
\caption{ $Hd(t)$ for the {\bf g} rule on the 6th structure. Size of
the system 1200.}
\end{figure}

 The dynamics lead by the rules {\bf b} and {\bf g} in the most extent
depends on the structure of interactions especially it is seen on the
second, third and 6th
structures, what allow to conclude that a spatial shape 
undoubtedly influence the dynamics of the system. But this influences is
essential only for the definite rules. The 6{\bf g} case has too large convergence
percentage for this rule.
In (Fig. 8) one can see the $Hd(t)$ distribution for the {\bf g} case of
the 6th structure.
The minimum of convergence percentage is observed in the {\bf c} case as well as for
the second structure.

\begin{figure}
\centerline{\epsfxsize11.5cm\epsffile{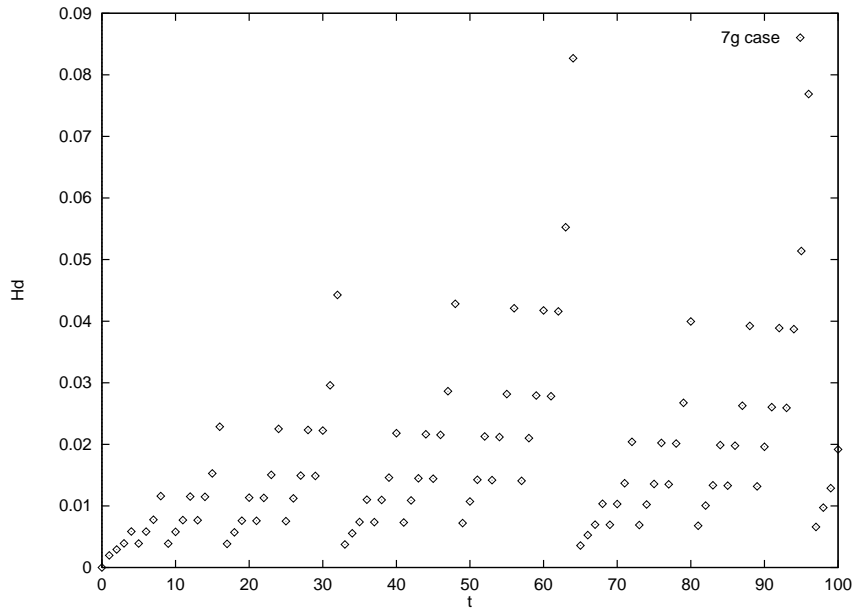}}
\caption{$Hd(t)$ for the {\bf g} rule on the 7th structure. Size of
the system 1020. }
\end{figure}

We studied the systems of different sizes.
It was obtained that an increase of the system's size does not give large
changes of dynamics, only slightly shifts percentage of
convergence in the direction of the present tendency, i.e. for rather order
cases it increases, for chaotic ones it decreases.

When we increase system's size, for the small ($<20$) maximum pattern differences the
results are the same and for large ones this value
increases. The
{\bf a} and {\bf g} rules have greater
divergence what results in a more branched transient structure.

Also we considered other rules with the same bias for all structures.

It was revealed that the {\bf a} and {\bf e} cases with the same bias give
results
independent
from the structure but strongly
dependent on the rules, behavior can change from chaos to order.
The maximum pattern difference is more stable than the convergence
percentage.
The cases with low convergence percentage are more stable to
rules' changes.
As for the rule {\bf g} it gives the most robust result on the quality
level.
The same rules'
changes result in different changes of system's dynamics for different
structures. 

\section{Conclusion}
\noindent
This paper represents the investigation of influence of a Boolean network
structure on behavior of the system. We have studied several different
structures, which differ by the geometric shape, lattice organization and
means of influence. We have used different rules of dynamics as well.
The means by which the investigation has been done consists of the measuring
of the averaged Hamming distance. In the work we used obtained for random
Boolean networks $K,p$ relation and compared the results with ones in 
random Boolean networks models resulted in dividing of system behavior into
ordered chaotic or marginal phases according to the $K,p$ relation. 

Influence of the spatial shape as well as the organization of
interactions on a lattice on system behavior is
quite confirmed by the results obtained.
As we saw both the second and the third structures form a cone
but system's dynamics in these cases has strong distinctions, the second
structure has 
the minimum average value of {\it mp}. 
At the same time the third
structure is characterized by the maximum value of the corresponding
parameter.

In the most extent the influence of the special shape of the network
has been observed for
the {\bf b} and {\bf g} rules.
For the second, third and sixth structures percentage of convergence
is essentially greater than for the other structures, that allows to
conclude that behavior of
Boolean Networks on a sphere is closer to a cone than to a torus
or a closed ribbon.

The closed lattices themselves do not influence on system's behavior but in
addition to hierarchy it decreases the convergence percentage, as it one
can see on example
of the 4,5,7 structures.

The influence of boundaries and adges is negligibly small.

Hierarchy in couple with irregularity gives the critical behavior
(2{\bf b},2{\bf d}) or shifts the character of system's dynamics in this
direction
(2{\bf g}).

The investigation of the second and the 7th structures pointed out that
there is no essential
influence of autoregulation.

$K,p$ relation obtained for random Boolean networks \cite{SS,C} has not been confirmed by the
results of our investigation of the concrete structuries.

In the larger number of cases small changes of the interaction's structure do not
give large changes of dynamics but in some cases, it is so.

{\bf Acknowledgment}.
\noindent
The work was partially supported by State Committee of Russian
Federation for high education grant No 97-14.3-58.


\newpage

\centerline{TABLES}

\begin{table}[htbp]
\caption{ In this table the data of convergence percentage of the couple of
trajectories in the phase space are presented. Sizes of the systems have order $10^3$
elements.}
\vskip0.3cm
\begin{center}
\begin{tabular}{|r|c|c|c|c|c|c|}
\hline
\tt  & a & b & c & d & e & g
\\
\hline
\tt 1 & 100 & 99.49  & 0.0441 & 99.54 & 99.56 & 0.00076
\\
\hline
\tt 2 & 100 & 49.95 & 0.08636 & 50.06 & 100 & 35.53
\\
\hline
\tt 3 & 100  & 32.8 & 66.4 & 100 & 99.997 & 36.17
\\
\hline
\tt 4 & 100 & 0.0765 & 0.428 &99.9992 & 99.9976 & 0.00366
\\
\hline
\tt 5 & 100 & 0.08937 & 0.4247 & 100& 99.9626 & 0.008
\\
\hline
\tt 6 & 100 & 13.995 & 0.413 & 100 & 99.982 & 80.053
\\
\hline
\tt 7 & 100 & 0.072 & 0.42 & 0.455 & 99.9973 & 0.0035
\\
\hline
\end{tabular}
\end{center}
\end{table}

\begin{table}[htbp]
\caption{ In this table the data of the maximum pattern difference between the couple of
simultaneously considered trajectories are
presented. Sizes of the systems have order $10^3$ elements.}
\vskip0.3cm
\begin{center}
\begin{tabular}{|r|c|c|c|c|c|c|}
\hline
\tt  & a & b & c & d & e & g
\\
\hline
\tt 1 & 500& 4 & 3 & 508 & 6 & 97
\\
\hline
\tt 2 & 1 & 10 & 2 & 3 & 1 & 7
\\
\hline
\tt 3 & 842 & 6 & 1023 & 16 & 15 & 292
\\
\hline
\tt 4 &  510& 4 & 10 & 7 & 11 & 49
\\
\hline
\tt 5 & 375 & 2 & 6 & 6& 10 & 58
\\
\hline
\tt 6 & 230 & 9 & 7 & 11 & 11 & 84
\\
\hline
\tt 7 & 495 & 3 & 8 & 18 & 12 & 208
\\
\hline
\end{tabular}
\end{center}
\end{table}

\begin{table}[htbp]
\caption{ In this table the data of period of $Hd(t)$ function are presented.
Sizes of the systems have order $10^3$ elements.
(pt $\equiv$ point, in other words the moment (time step) when all envisaged pairs of
trajectores have converged)}
\vskip0.3cm
\begin{center}
\begin{tabular}{|r|c|c|c|c|c|c|}
\hline
\tt  & a & b & c & d & e & g
\\
\hline
\tt 1 & 0 from 256 pt & 2 & 1 &$ >50$ & 1&$ >50$
\\
\hline
\tt 2 & 0 from 9 pt & 1 & 2 & 2 & 0 from 11 pt & 8
\\
\hline
\tt 3 & 0 from 16 pt & 6 & 12 & 0 from 21 pt & 1 &$>50$
\\
\hline
\tt 4 & 0 from 101 pt & 1 & 1 & 0 from 18 pt & 2 &$ >50$
\\
\hline
\tt 5 & 0 from 148 pt & 1 & 1 & 0 from 13 pt & 2 &$ >50$
\\
\hline
\tt 6 & 0 from 39 pt & 8 & 1 & 0 from 34 pt & 2 & $>50$
\\
\hline
\tt 7 & 0 from 105 pt & 1 & 4 & 4 & 3 & $>50 $
\\
\hline
\end{tabular}
\end{center}
\end{table}

\end{document}